\documentclass{ws-procs9x6}

\begin{document}

\title{On coherent radiation in electron-positron colliders
\footnote{\uppercase{T}his work is partially 
supported by grant 03-02-16154 of the \uppercase{R}ussian 
\uppercase{F}und of \uppercase{F}undamental \uppercase{R}erearch.}}

\author{V.~N. BAIER and V.~M.KATKOV}

\address{Budker Institute of Nuclear Physics, \\
Novosibirsk, 630090, Russia\\ 
E-mail: baier@inp.nsk.su; katkov@inp.nsk.su}




\maketitle

\abstracts{
The electromagnetic processes in linear colliders are discussed on the basis of  
quasiclassical operator method. The complete set of expressions 
is written down for spectral probability of radiation from an electron 
and pair creation by a photon taking into account both electron 
and photon polarization.
Some new formulas are derived
for radiation intensity and its asymptotics. The main mechanisms
of pair creation dominate at $\chi \leq 1$ are discussed.}

\section{Introduction}

The particle interaction at beam-beam collision in linear colliders 
occurs in an electromagnetic
field provided by the beams. As a result, 
1)the phenomena induced by this field turns out to be 
very essential, 2)the cross section of the main QED processes are 
modified comparing to the case of free particles. These items were considered 
by V.M.Strahkhovenko and authors \cite{BKS1},
\cite{BKS2}.

The magnetic bremsstrahlung mechanism dominates and its characteristics are 
determined by the 
value of the quantum parameter $\chi(t)$ dependent on the 
strength of the incoming beam
field at the moment $t$ (the constant field limit)
\begin{equation}
\chi^2=-\frac{e^2}{m^2}(F_{\mu\nu}p_\nu)^2, \quad \chi=\frac{\gamma F}{H_0},
\label{1.1}
\end{equation} 
where $p^\nu(\varepsilon,\textbf{p})$ is a particle four-momentum, $F^{\mu\nu}$ 
is an external
electromagnetic field tensor, $\gamma=\varepsilon/m$, 
$\textbf{F}=\textbf{E}_\bot+\textbf{v}
\times\textbf{H}$, $\textbf{E}$ and $\textbf{H}$ are the electric 
and magnetic fields in the laboratory
frame, $\textbf{E}_\bot=\textbf{E}-\textbf{v}(\textbf{vE})$, \textbf{v} 
is the particle velocity, $F=|\textbf{F}|$ and
$H_0=m^2/e=(m^2c^3/e\hbar)=4.41\cdot 10^{13}$~Oe. We employ units $\hbar=c=1$
and $\alpha=1/137$, 

\subsection{General formulas}

The photon radiation length in an external field is
\begin{equation}
l_c(\chi, u) = \lambda_c \frac{H_0}{F}\left(1+\frac{\chi}{u} \right)^{1/3} 
= \frac{\lambda_c \gamma}{\chi}\left(1+\frac{\chi}{u} \right)^{1/3},
\label{2.1}
\end{equation} 
where $\lambda_c=\hbar/mc$ is the electron Compton wave length, 
 $u=\omega/\varepsilon',
\varepsilon'=\varepsilon-\omega$, $\omega$ is the photon energy. 
The field of the incoming beam changes very slightly 
along the formation length $l_c$, if the condition $l_c \ll \sigma_z$ ($\sigma_z$ 
is the
longitudinal beam size) is satisfied, providing a high accuracy of the magnetic 
bremsstrahlung approximation. 

In the general case, when both polarizations of electron and photon are taken
into account, the spectral probability of radiation from an electron
per unit time has the form\cite{BKF1}
(see also \cite{BKS2})
\begin{eqnarray}
\hspace{-7mm}&& \frac{dw_{\gamma}}{dt} \equiv dW_{\gamma}(t)=
\frac{\alpha}{2\sqrt{3}\pi \gamma^2}
\Phi_{\gamma}^{\zeta}(1+(\mbox{\boldmath$\lambda$}\mbox{\boldmath$\xi$}))d\omega;~
\Phi_{\gamma}^{\zeta}=\Phi_{\gamma}-\frac{\omega}{\varepsilon}
(\mbox{\boldmath$\zeta$}{\bf h})K_{1/3}(z),
\nonumber \\
\hspace{-7mm}&&\Phi_{\gamma}(t)=\left(\frac{\varepsilon}{\varepsilon'}+
\frac{\varepsilon'}{\varepsilon} \right)K_{2/3}(z)-\int_{z}^{\infty}K_{1/3}(y)dy,
\label{1} 
\end{eqnarray} 
where $K_{\nu}(z)$ is the Macdonald functions, $z=2u/3\chi(t)$,
$\mbox{\boldmath$\lambda$}(\lambda_1,\lambda_2,\lambda_3)$ are
the Stokes parameters of emitted photons for the following choice of axes:
${\bf e}_1=({\bf v}\times{\bf h}),~{\bf h}={\bf F}^{\ast}/F,~{\bf e}_2={\bf h},~
{\bf F}^{\ast}=e/|e|[{\bf H}_{\perp}+({\bf E}\times{\bf v})]$, $e$ is 
the charge of particle, $\mbox{\boldmath$\zeta$}$ is the spin vector of 
the initial electron in its rest frame. The vector $\mbox{\boldmath$\xi$}$
determines the mean photon polarization and its 
components are given by the following expressions:
\begin{eqnarray}
\hspace{-7mm}&& \xi_1=
\frac{\omega (\mbox{\boldmath$\zeta$}{\bf v}{\bf h})}{\varepsilon'\Phi_{\gamma}^{\zeta}}
K_{1/3}(z),\quad \xi_3=\frac{1}{\Phi_{\gamma}^{\zeta}}
\left[K_{2/3}(z)+\frac{\omega}{\varepsilon'}(\mbox{\boldmath$\zeta$}{\bf h})
K_{1/3}(z) \right], 
\nonumber \\
\hspace{-7mm}&& \xi_2=\frac{(\mbox{\boldmath$\zeta$}{\bf v})}{\Phi_{\gamma}^{\zeta}}
\Bigg[ \left(\frac{\varepsilon}{\varepsilon'}-
\frac{\varepsilon'}{\varepsilon} \right)K_{2/3}(z)
 -\frac{\omega}{\varepsilon}\int_{z}^{\infty}K_{1/3}(y)dy\Bigg],
\label{2} 
\end{eqnarray} 
here $(\mbox{\boldmath$\zeta$}{\bf v}{\bf h})
=\mbox{\boldmath$\zeta$}({\bf v}\times{\bf h})$.

Using asymptotic expansion of the Macdonald functions 
$K_{\nu}(z)=(\Gamma(\nu)/2)(2/z)^{\nu},~(z \ll 1)$ we obtain for
the case $u \ll \chi$
\begin{eqnarray}
 &&\Phi_{\gamma}^{\zeta}=\Gamma\left(\frac{2}{3}\right)
\left( \frac{3\chi}{u}\right)^{2/3} \left(1+
\frac{\omega^2}{2\varepsilon\varepsilon'} \right);
\nonumber \\
&&\xi_1=0,~
\xi_2=\frac{\varepsilon^2-\varepsilon'^2}{\varepsilon^2+\varepsilon'^2}
(\mbox{\boldmath$\zeta$}{\bf v}),~\xi_3
=\frac{\varepsilon\varepsilon'}{\varepsilon^2+\varepsilon'^2}.   
\label{3}
\end{eqnarray} 
It is seen from given here characteristics
that, generally speaking, the radiation is polarized. 
For unpolarized initial electrons $\xi_3 \neq 0$.

The probability of pair creation by a photon in the external field can
be find from formulas (\ref{1}), (\ref{2}) using the substitution rule \cite{BKF1}:
$\varepsilon \rightarrow -\varepsilon, \quad \omega \rightarrow -\omega,
\quad \mbox{\boldmath$\zeta$} \rightarrow -\mbox{\boldmath$\zeta$},\quad
\lambda_2 \rightarrow -\lambda_2,
\quad \lambda_{1,3} \rightarrow \lambda_{1,3}({\bf e} \rightarrow {\bf e}^{\ast}),\quad
\omega^2d\omega \rightarrow - \varepsilon^2d\varepsilon$. 
Performing these substitutions we obtain
\begin{eqnarray}
\hspace{-7mm}&& \frac{dw_{e}}{dt} \equiv dW_{e}(t)=\frac{\alpha m^2}{2\sqrt{3}\pi \omega^2}
\Phi_{e}^{\zeta}(1+(\mbox{\boldmath$\lambda$}\mbox{\boldmath$\Sigma$}))d\varepsilon;~
\Phi_{e}^{\zeta}=\Phi_{e}-\frac{\omega}{\varepsilon}
(\mbox{\boldmath$\zeta$}{\bf h})K_{1/3}(y),
\nonumber \\
\hspace{-7mm}&&\Phi_{e}(t)=\left(\frac{\varepsilon}{\varepsilon'}+
\frac{\varepsilon'}{\varepsilon} \right)K_{2/3}(y)+\int_{y}^{\infty}K_{1/3}(x)dx,
,~\Sigma_1=
-\frac{\omega (\mbox{\boldmath$\zeta$}{\bf v}{\bf h})}{\varepsilon'\Phi_{e}^{\zeta}}
K_{1/3}(y),
\nonumber \\
\hspace{-7mm}&& \Sigma_2=\frac{(\mbox{\boldmath$\zeta$}{\bf v})}{\Phi_{e}^{\zeta}}
\Bigg[ \left(\frac{\varepsilon}{\varepsilon'}-
\frac{\varepsilon'}{\varepsilon} \right)K_{2/3}(y) +
\frac{\omega}{\varepsilon}\int_{y}^{\infty}K_{1/3}(x)dx\Bigg],
\nonumber \\
\hspace{-7mm}&&\Sigma_3=-\frac{1}{\Phi_{e}^{\zeta}}
\left[K_{2/3}(y)+\frac{\omega}{\varepsilon'}(\mbox{\boldmath$\zeta$}{\bf h})
K_{1/3}(y) \right],~y=\frac{2\omega^2}{3\varepsilon\varepsilon'\kappa},~
\kappa=\frac{\omega F}{m H_0} 
\label{4} 
\end{eqnarray} 
here $\varepsilon$ is the energy of the created electron,
$\varepsilon'=\omega-\varepsilon$ is the energy of created positron,
$\mbox{\boldmath$\zeta$}$ is the electron spin vector, 
$\mbox{\boldmath$\lambda$}(\lambda_1, \lambda_2, \lambda_3)$ are the 
Stokes parameters of the initial photon. 
 
Integrating (\ref{4}) 
over $\varepsilon$ (in some terms, integration by parts was carried out)
we get the total probability of pair creation (per unit time) \cite{BKS3}
\begin{eqnarray}
\hspace{-7mm}&& W_e^{\zeta}=\frac{1}{2}\left(W_e-
\frac{\alpha m^2}{\sqrt{3}\pi \omega}(\mbox{\boldmath$\zeta$}{\bf h})
\int_{0}^{1} \frac{K_{1/3}(\eta)}{x}dx\right),
\nonumber \\
\hspace{-7mm}&& W_e=\frac{2\alpha m^2}{3\sqrt{3}\pi \omega}
\int_{0}^{1} K_{2/3}(\eta)\Bigg[\frac{1-3\lambda_3}{2}
+\frac{1}{x(1-x)} \Bigg]dx,
\label{5} 
\end{eqnarray} 
where $x=\varepsilon/\omega$ and $\eta=2/(3\kappa x(1-x))$.

For longitudinally polarized initial electrons 
(see Eqs.(\ref{2}),(\ref{3})) the hard photons 
($\omega \simeq \varepsilon,~\varepsilon' \ll \varepsilon$) 
are circularly polarized. The polarization
of created electrons and positrons is discussed in detail 
in\cite{BKS4b}. In particular, for the circular polarization of
incoming photon the created electrons with $x \rightarrow 1$ 
have the longitudinal polarization (see (\ref{4})). 
All these effects are manifestation of
"the helicity transfer". 

\section{Photon Emission}

Here we consider radiation from unpolarized electrons.
The spectral probability of radiation is (\ref{1})
\begin{equation}
\frac{dw_{\gamma}}{d\omega}=
\frac{\alpha}{\pi\gamma^2 \sqrt{3}}\int_{-\infty}^{\infty}
\Phi_{\gamma}(t)dt.
\label{2.2} 
\end{equation} 
For the Gaussian beams
\begin{equation}
\chi(t)=\chi_0(x,y)\exp(-2t^2/\sigma_z^2),
\label{2.3} 
\end{equation} 
here the function $\chi_0(x,y)$ depends on transverse coordinates.

It turns out that for the Gaussian beams the integration of 
the spectral probability over time can be carried 
out in a general form:
\begin{eqnarray}
&& \frac{dw_{\gamma}}{du}=\frac{\alpha m \sigma_z}{\pi\gamma \sqrt{6}}
\frac{1}{(1+u)^2}\Bigg[\left(1+u+\frac{1}{1+u} \right) 
\nonumber \\
&&\times \int_{1}^{\infty}K_{2/3}\left(ay \right) 
\frac{dy}{y\sqrt{\ln y}}-2a\int_{1}^{\infty}K_{1/3}\left(ay \right) 
\sqrt{\ln y}~dy\Bigg], 
\label{2.4}
\end{eqnarray} 
where $a=2u/3\chi_0$. 
In the case when $a \gg 1$ the main contribution into integral (\ref{2.4})
gives the region $y=1+\xi, \xi \ll 1$. Taking the integrals over $\xi$ we obtain
\begin{equation}
\frac{dw_{\gamma}^{CF}}{du} \simeq \frac{\sqrt{3}\alpha m \sigma_z}{4\gamma}
\frac{1+u+u^2}{u(1+u)^3} \chi_0 \exp\left( -\frac{2u}{3\chi_0}\right) 
\label{2.5}
\end{equation} 

For round beams the integration over transverse coordinates is 
performed with the density
\begin{equation}
n_{\perp}(\mbox{\boldmath$\varrho$})=\frac{1}{2\pi \sigma_{\perp}^2}
\exp\left(-\frac{\varrho^2}{2\sigma_{\perp}^2} \right) 
\label{2.6}
\end{equation} 
The parameter $\chi_0(\varrho)$ we present in the form
\begin{eqnarray}
&&\chi_0(\varrho)=\chi_m\frac{f(x)}{f_0},\quad x=\frac{\varrho}{\sigma_{\perp}},
\quad f(x)=\frac{1}{x}\left(1-\exp(-x^2/2) \right),
\nonumber \\
&&\chi_{rd}=0.720 \alpha N \gamma \frac{\lambda_c^2}{\sigma_z\sigma_{\perp}},~
 f'(x_0)=0,~f_0=f(x_0)=0.451256,
\label{2.7}
\end{eqnarray} 
where $N$ is the number of electron in the bunch.

\begin{figure}[ht]
\centerline{\epsfxsize=3.9in\epsfbox{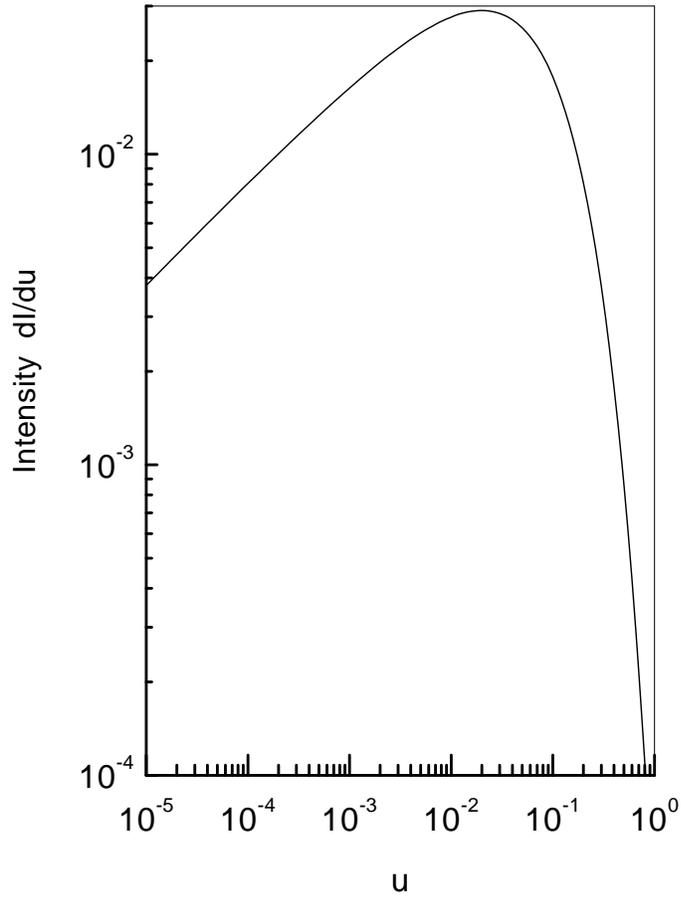}}   
\caption{Spectral intensity of radiation of round beams in units 
 $\alpha m^2 \sigma_z$ for 
 $\chi_{rd}$=0.13 calculated according to Eqs.(\ref{2.4}),(\ref{2.7}). \label{Fig.1}}
\end{figure}

The Laplace integration of Eq.(\ref{2.5}) gives for radiation intensity 
$dI/du=\varepsilon u/(1+u)dW/du$
\begin{equation}
\frac{dI_{as}}{du} \simeq \alpha m^2 \sigma_z 
\frac{3}{4}\sqrt{\frac{\pi}{|f''_0|}}
\frac{1+u+u^2}{\sqrt{u}(1+u)^4}
 f_0^{3/2}\chi_{rd}^{3/2}\exp\left(-\frac{2u}{3\chi_{rd}}\right),
\label{2.8} 
\end{equation}  
where $f''_0=f''(x_0)=-0.271678$. 

Integration of (\ref{2.4}) 
over transverse coordinates gives
the final result for the radiation intensity. 
For the round beams it is shown in Fig.1 for $\chi_{rd}=0.13$,
the curve attains the maximum at $u \simeq 0.02$. The right slope 
of the curve agrees with the asymptotic 
intensity (\ref{2.8}) and the left slope of the curve agrees 
with the standard classical intensity. 
\begin{equation}
\frac{dI_{cl}}{du}=\frac{e^2 m^2}{\pi}3^{1/6}\Gamma(2/3)\chi^{2/3}u^{1/3}
\label{2.8a}
\end{equation}

It will be instructive to compare the spectrum in Fig.1, found by means of
integration over the transverse coordinates with intensity spectrum which
follows from Eq.(\ref{2.4}) (multiplied by $\omega$) with averaged 
over the density Eq.(\ref{2.6}) value $\chi_0$ Eq.(\ref{2.7}):
$\overline{\chi_0}=\chi_{rd} \cdot 0.8135 $.
The last spectrum reproduces the spectrum given in Fig.1,
in the interval $10^{-3} \leq u/\overline{\chi_0} \leq 1$ with an accuracy
better than 2\% (near maximum better than 1\%) 
while for $ u/\overline{\chi_0} \leq 10^{-3}$ one can use the classic intensity
(\ref{2.8a}) and for $ u/\overline{\chi_0} \geq 1$ the asymptotics (\ref{2.8})
is applicable.

For the flat beams ($\sigma_x \gg \sigma_y$) 
the parameter $\mbox{\boldmath$\chi$}_0(\mbox{\boldmath$\varrho$})$ takes the form
\begin{equation}
\mbox{\boldmath$\chi$}_0=\frac{2\gamma {\bf E}_{\perp}}{H_0}=  
\chi_m e^{-v^2}
\left[\textbf{e}_y{\rm erf}\left(w \right) 
 -i \textbf{e}_x{\rm erf}\left(iv \right) \right],\quad
 \chi_m= \frac{2N\alpha\gamma \lambda_c^2}{\sigma_z\sigma_x},
\label{2.9} 
\end{equation} 
here $v=x/\sqrt{2}\sigma_x,~w=y/\sqrt{2}\sigma_y$, 
${\rm erf}(z)=2/\sqrt{\pi}\int_{0}^{z}\exp(-t^2)dt$, 
$\textbf{e}_x$ and $\textbf{e}_y$
are the unit vectors along the corresponding axes. The formula (\ref{2.9}) 
is consistent with given in \cite{N}.  In \cite{BKS1},\cite{BKS2} 
the term with $\textbf{e}_x$
was missed. Because of this the numerical coefficients in results for the flat beams are 
erroneous.

To calculate the asymptotics of radiation intensity for the case $u \gg \chi_m$ one has 
to substitute
\begin{equation}
\chi_0=|\mbox{\boldmath$\chi$}_0|=\chi_m
e^{-v^2}   
\left[ {\rm erf}^2\left(w \right)
-{\rm erf}^2\left(iv \right)
\right]^{1/2} 
\label{2.10}
\end{equation} 
into Eq.(\ref{2.5}) and take integrals over transverse coordinates $x,y$ with the weight
\begin{equation}
n_{\perp}(x,y)=\frac{1}{2\pi \sigma_x \sigma_y}\exp\left(-\frac{x^2}{2\sigma_x^2} 
-\frac{y^2}{2\sigma_y^2} \right).
\label{2.11}
\end{equation} 
Integral over $x$ can be taken using the Laplace method, while for integration over $y$
it is convenient to introduce the variable 
\begin{equation}
\eta=\frac{2}{\sqrt{\pi}}\int_{w}^{\infty}\exp(-t^2)dt,\quad w=\frac{y}{\sqrt{2}\sigma_y}.
\label{2.12}
\end{equation} 
As a result we obtain for the radiation intensity in the case of flat beams
\begin{equation}
\frac{dI_{fl}}{du}=\frac{9}{8\sqrt{2\left(1-2/\pi \right)}} \alpha m^2 \sigma_z
\chi_m^{5/2} 
 \frac{1+u+u^2}{u^{3/2}(1+u)^4}\exp\left( -\frac{2u}{3\chi_m}\right).
\label{2.13}
\end{equation} 
It is interesting to compare the high-energy end of intensity 
spectrum at collision of flat beams (\ref{2.13}) with intensity spectrum
of incoherent radiation with regard for smallness of the transverse beam 
sizes \cite{BK1}.
For calculation we use the project TESLA parameters \cite{T}:
~$\varepsilon=250$~GeV,~$\sigma_x=553$~nm, 
$\sigma_y=5$~nm,~$\sigma_z=0.3$~mm, $N=2\cdot 10^{10}$, $\chi_m$=0.13. 
The result is shown in Fig.2, where the curves 
calculated according to Eq.(\ref{2.13}),
and Eq.(5.7) in \cite{BK1}. 
It is seen that for $x=\omega/\varepsilon > 0.7$ 
the incoherent radiation dominates. For $x > 0.7$ the incoherent radiation
may be used for a tuning of beams \cite{BK1}.

\begin{figure}[ht]
\centerline{\epsfxsize=3.9in\epsfbox{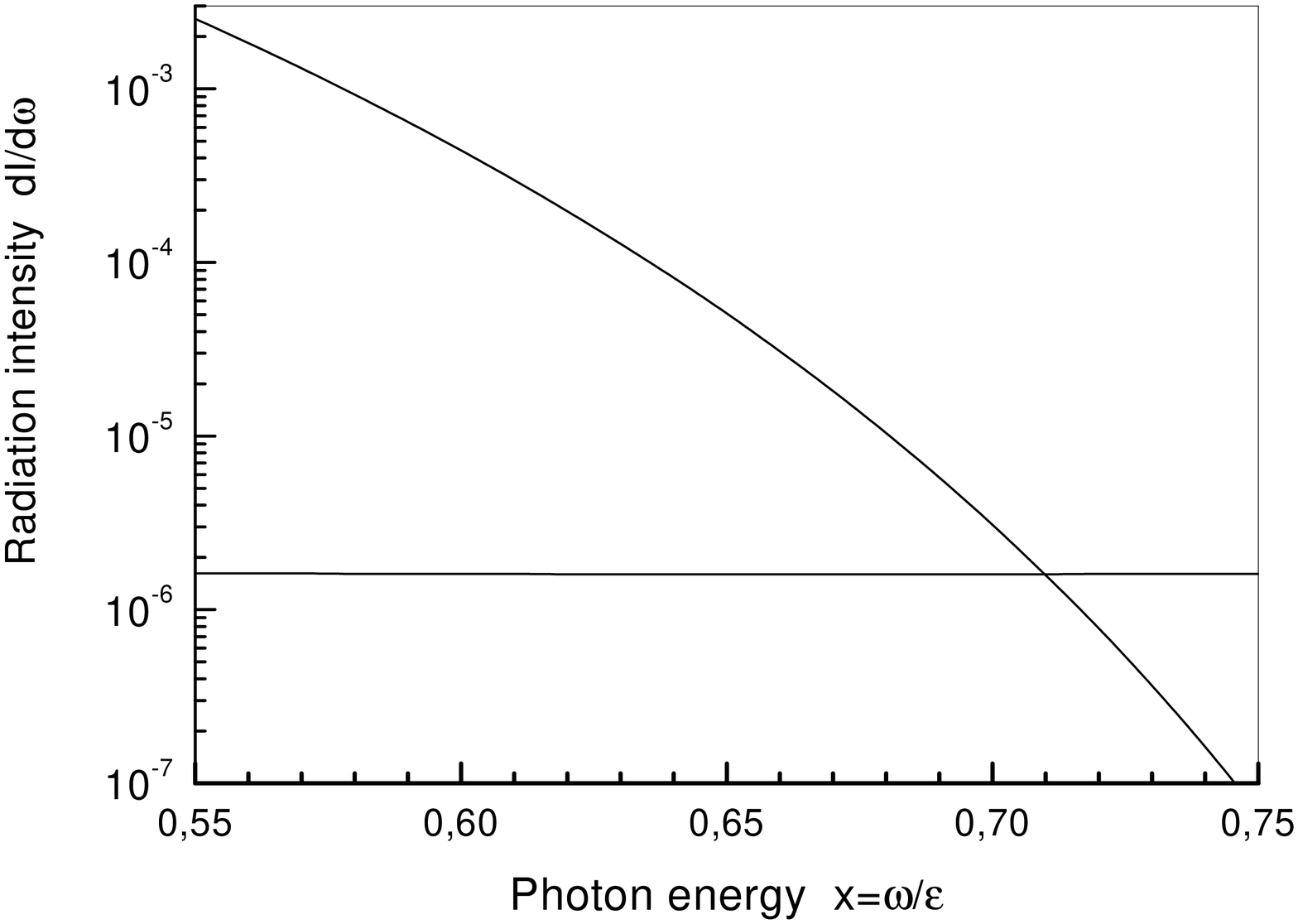}}   
\caption{The spectral radiation intensity $dI/d\omega$ 
of coherent radiation (fast falling with 
$x=\omega/\varepsilon$ increase curve) and 
of incoherent radiation (the curve which is almost
constant) in units $N\alpha^2\lambda_c/\sigma_x$ 
for beams with dimensions $\sigma_x=553$~nm, 
$\sigma_y=5$~nm, for $\chi_m$=0.13. \label{Fig.2}}
\end{figure}

Along with the spectral characteristics of radiation the total
number of emitted by an electron photons is of evident interest as
well as the relative energy losses. We discuss an actual case of 
flat beams and the parameter $\chi_m \ll 1$. In this case one can use
the classic expression for intensity (bearing in mind that at  
$\chi_m \geq 1/10$ the quantum effects become substantial).
In classical limit the relative energy losses are
\begin{equation}
\frac{\Delta \varepsilon}{\varepsilon}=\frac{2\alpha m^2}{3\varepsilon}
\int \chi^2(t, x, y) n_{\perp}(x,y)dt dx dy.
\label{2.14}
\end{equation} 
Using Eqs.(\ref{2.3}), (\ref{2.10}), and (\ref{2.11}) we get
\begin{equation}
\frac{\Delta \varepsilon}{\varepsilon}=\frac{2}{9}\sqrt{\frac{\pi}{3}}r \chi_m^2,
\quad r=\frac{\alpha \sigma_z}{\gamma \lambda_c}.
\label{2.15}
\end{equation}
For mean number of emitted by an electron photons we find
\begin{equation}
n_{\gamma}=\frac{5\alpha m^2}{2\sqrt{3}\varepsilon}
\int \chi(t, x, y) n_{\perp}(x,y)dt dx dy
=0.59275 \frac{5}{2}\sqrt{\frac{\pi}{6}}r \chi_m
=1.072 r \chi_m. 
\label{2.16}
\end{equation}
If $\chi_m > 1/10$ one have to use the quantum formulas. For 
the energy losses one can use the  approximate expression (the accuracy
is better than 2\% for any $\chi$) \cite{BKS5}
\begin{equation}
d\varepsilon/dt=2/3 \alpha m^2 \chi^2
\left[1+4.8(1+\chi)\ln(1+1.7\chi+2.44\chi^2) \right]^{-2/3}.
\label{2.17}
\end{equation}
Here $\chi$ is the local value, so this expression for 
$d\varepsilon/dt$ should be integrated over time and averaged over the 
transverse coordinates.
For mean number of photons emitted by an electron there is the 
approximate expression (the accuracy
is better than 1\% for any $\chi_0$)\cite{BKS2}
\begin{equation}
n_{\gamma}(\mbox{\boldmath$\varrho$})
=\frac{1.81\chi_0 r}{[1+1.5(1+\chi_0)\ln(1+3\chi_0)+0.3\chi_0^2]^{1/6}}
\label{2.18},
\end{equation}
where the expression should be averaged over the transverse coordinates.
For the project TESLA one gets for 
$(\Delta \varepsilon/\varepsilon)_{cl}=4.3\%$ according to (\ref{2.15}),
while the correct result from (\ref{2.17}) is 
$(\Delta \varepsilon/\varepsilon)_{cl}=3.2\%$.
For mean number of emitted by an electron photons we have correspondingly
$n_{\gamma}^{cl} \simeq 1.6$ while the correct result is
$n_{\gamma} \simeq 1.5$.

\section{Pair Creation}

There are different mechanisms of electron-positron pair creation
\begin{enumerate}
\item Real photon radiation in the field and pair creation 
by this photon in
the same field of the opposite beam. This mechanism dominates 
in the case $\chi \geq 1$.
\item Direct electroproduction of electro-positron pair in the 
field through virtual photon.
This mechanism is also essential in the case $\chi \geq 1$.
\item Mixed mechanism(1):photon is radiated in the bremsstrahlung 
process, i.e. incoherently
and the pair is produced in external field.
\item Mixed mechanism(2): photon is radiated in a magnetic 
bremsstrahlung way, 
and pair is produced in the interaction of this photon with 
individual particles of 
oncoming beam,
i.e. in interaction with potential fluctuations.
\item Incoherent electroproduction of pair.
\end{enumerate} 
All these types of pair production processes were considered 
in detail in \cite{BKS2}.
In the actual case $\chi \ll 1$ mixed and incoherent 
mechanisms mostly contribute. 
We start with the mixed mechanism (2). In the case 
$\chi_m \ll 1$ the parameter $\kappa$ Eq.(\ref{4}) containing
the energy of emitted photon is also small and the incoherent cross
section of pair creation by a photon is weakly dependent on the
photon energy
\begin{equation}
\sigma_p=28/9\alpha^3 \lambda_c^2 \ln(\sigma_y/\lambda_c)
\left(1+ 396\kappa^2/1225 \right).
\label{3.1}
\end{equation}
If the term $\propto\kappa^2 $ being neglected, the pair creation 
probability is factorized (we discuss coaxial beams of 
identical configuration). The number of pairs created by this mechanism  
(per one initial electron) is
\begin{eqnarray}
\hspace{-7mm}&&n_p^{(2)}=\int_{-\infty}^{\infty}
W_{\gamma}(\mbox{\boldmath$\varrho$}, t)dt
\int_{t}^{\infty}2\sigma_p n_{\perp}(\mbox{\boldmath$\varrho$}, t')dt'd^2\varrho
\nonumber \\
\hspace{-7mm}&&=\frac{35}{9}\sqrt{\frac{\pi}{6}}N r 
\alpha^3 \lambda_c^2 \ln \frac{\sigma_y}{\lambda_c}
\int \chi_0(\mbox{\boldmath$\varrho$}) 
n_{\perp}^2 (\mbox{\boldmath$\varrho$}d^2\varrho)
=\frac{35}{9}\sqrt{\frac{\pi}{6}}N r \chi_m \alpha^3 
\frac{\lambda_c^2}{2\pi\sigma_x\sigma_y }\frac{4}{\pi}
\nonumber \\
\hspace{-7mm}&&
\times \int_{0}^{\infty}\exp(-3v^2-2w^2) 
\big[  {\rm erf}^2\left(w \right)-{\rm erf}^2\left(iv \right)
\big]^{1/2} dw dv 
\nonumber \\
\hspace{-7mm}&&=0.1167\alpha^3 N r \chi_m \frac{\lambda_c^2}{\sigma_x\sigma_y }
\ln \frac{\sigma_y}{\lambda_c}.
\label{3.2}
\end{eqnarray}
So for the project  TESLA parameters the total number of produced pairs 
by this mechanism per bunch in one collision  
is $Nn_p^{(2)} \sim 1.4 \cdot 10^4$.

Now we turn over to discussion of incoherent electroproduction of pairs when
both intermediate photons are virtual. To within the logarithmic accuracy
for any $\chi$ and $\kappa$ 
one can use the method of equivalent photons
\begin{equation}
\frac{d\sigma_v}{d\omega}=n(\omega)\sigma_p(\omega),
\label{3.7}
\end{equation} 
where for soft photons $(\omega \ll \varepsilon)$
\begin{eqnarray}
\hspace{-7mm}&& n(\omega)=\frac{2\alpha}{\pi}\ln\frac{\Delta}{q_m},
~\Delta=m(1+\kappa)^{1/3},
\nonumber \\
\hspace{-7mm}&& q_m=m\frac{\omega}{\varepsilon}\left(1+
\frac{\varepsilon \chi}{\omega}\right)^{1/3} +
\frac{1}{\sigma_y}\equiv q_F +q_\sigma.
\label{3.8}
\end{eqnarray} 
Taking into account that the cross section of pair photoproduction is
\begin{equation}
\sigma_p(\omega)=\frac{28 \alpha^3}{9m^2}\ln\frac{m}{q_m(\omega')},~
\omega'=\frac{m^2}{\omega}~(\kappa \ll 1);~\sigma_p \propto \kappa^{-2/3}~
(\kappa \gg 1),
\label{3.9}
\end{equation}
we obtain in the main logarithmic approximation for the cross section of the 
pair electroproduction 
\begin{equation}
\sigma(2e \rightarrow 4e)= \frac{56 \alpha^4}{9m^2}
\int_{\omega_{min}}^{\omega_{max}}
\ln\frac{m}{q_m(\omega)}\ln\frac{m}{q_m(\omega')}\frac{d\omega}{\omega},
\label{3.10}
\end{equation} 
where $\omega_{max}=\varepsilon/(1+\chi),~\omega_{min}=m^2/\omega_{max}$.
If we put $\chi=0,~\sigma_y=\infty$ we obtain the standard Landau-Lifshitz
cross section $\sigma_{LL}$ (see e.g. \cite{BKF1})
\begin{equation}
\sigma_{LL}= \frac{28 \alpha^4}{27m^2}\ln^3\gamma^2.
\label{3.11}
\end{equation} 
With regard for the bounded transverse dimensions of beam and
influence of an external field the equivalent photon spectrum 
changes substantially. For $\chi \sim 1$ we have
$q(\omega)=m(\omega/\varepsilon)^{2/3}+1/\sigma_y$ and under
condition $\gamma^{2/3}\lambda_c/\sigma_y \geq 1$ we find
\begin{equation}
\sigma_{v}= \frac{28 \alpha^4}{3m^2}\ln\left(\gamma^{4/3}
\frac{\lambda_c}{\sigma_y}\right)\ln^2\frac{\sigma_y}{\lambda_c}.
\label{3.12}
\end{equation}
For TESLA parameters $\gamma^{2/3}\lambda_c/\sigma_y \simeq 1$
then 
\begin{equation}
\sigma_{v}= \frac{28 \alpha^4}{81m^2}\ln^3\gamma^2.
\label{3.13}
\end{equation}
This cross section is three times smaller than standard $\sigma_{LL}$.
For the project TESLA parameters the number of pairs 
produced by this mechanism per bunch in one collision 
$n_v= L\sigma_{v} \simeq 1.5 \cdot 10^4,~L=1/(4\pi \sigma_x \sigma_y )$ is 
the geometrical luminosity per bunch\cite{BK1}.
So the both discussed mechanisms give nearly the same contribution for this
project. 

It should be noted that the above analysis was performed under assumption
that the configuration of beams doesn't changed during collision, although
in the TESLA project the disruption parameter $D_y > 1$.


\begin{thebibliography}{99}

\bibitem{BKS1} V.N.Baier, V.M.Katkov,and  V.M.Strakhovenko,
{\it Particle Accelerators}, {\bf 30},  43 (1990).

\bibitem{BKS2} V.N.Baier, V.M.Katkov, and V.M.Strakhovenko,
{\it Preprint INP 89-180}, Novosibirsk 1989 (unpublished).

\bibitem{BKF1} V.N.Baier, V.M.Katkov,and V.S.Fadin,
{\it Radiation from Relativistic Electrons}, Atomizdat, 
Moscow, 1973 (in Russian).

\bibitem{BKS3} V.N.Baier, V.M.Katkov, and  V.M.Strakhovenko,
{\it Phys. Lett}. {\bf  B229}, 135 (1989).

\bibitem{BKS4b} V.N.Baier, V.M.Katkov, and  V.M.Strakhovenko,
{\em Electromagnetic Processes at High Energies in Oriented
Single Crystals}, World Scientific, Singapore, 1998.

\bibitem{N} R.J.Noble, {\it Nucl.Instr.Meth.},
{\bf A256}, 427 (1987).

\bibitem{BKS31} V.N.Baier, V.M.Katkov, and  V.M.Strakhovenko,
{\it Sov.Phys.JETP}, {\bf 67}, 70 (1988).

\bibitem{T} O.Napoly, 
{\it Proceedings of the 2001 Particle Accelerator Conference, Chicago}, 
p.402, 2001.

\bibitem{BKS5} V.N.Baier, V.M.Katkov, and  V.M.Strakhovenko,
{\it Phys. Lett}. {\bf  A229}, 429 (1992).

\bibitem{BKS4} V.N.Baier, V.M.Katkov,and  V.M.Strakhovenko,
{\it Sov.J.Nucl.Phys}. {\bf 36}, 95 (1982).

\bibitem{BK1} V.N.Baier, V.M.Katkov
{\it Phys. Rev}, {\bf D66},  053009 (2002).

\end{thebibliography}
\end{document}